\documentclass[sigconf]{acmart}
\usepackage{listings}
\usepackage{algpseudocode}
\usepackage{algorithm}
\newfloat{lstfloat}{htbp}{lop}
\floatname{lstfloat}{Listing}
\setcopyright{none}
\renewcommand\footnotetextcopyrightpermission[1]{}
\usepackage{footnote}
\usepackage{xcolor}
\usepackage{tikz}

\lstdefinelanguage{mlir}{
    alsodigit = {.},
    keywords = {stencil.apply, stencil.access, arith.constant, arith.addf, arith.mulf, stencil.return, f32, f64, stencil.index,stencil.temp}
}

\AtBeginDocument{%
  \providecommand\BibTeX{{%
    \normalfont B\kern-0.5em{\scshape i\kern-0.25em b}\kern-0.8em\TeX}}}

\newcommand\copyrighttext{%
  \footnotesize \textcopyright Nick Brown 2023. This is the author's version of the work. It is posted here for your personal use. Not for redistribution. The definitive version was published in ACM Workshops of The International Conference on High Performance Computing, Network, Storage, and Analysis (SC-W 2023), https://doi.org/10.1145/3624062.3624167"}
\newcommand\copyrightnotice{%
\begin{tikzpicture}[remember picture,overlay]
\node[anchor=south,yshift=10pt] at (current page.south) {\fbox{\parbox{\dimexpr\textwidth-\fboxsep-\fboxrule\relax}{\copyrighttext}}};
\end{tikzpicture}%
}

\begin{document}

\title{Fortran performance optimisation and auto-parallelisation by leveraging MLIR-based domain specific abstractions in Flang}

\author{Nick Brown}
\email{n.brown@ed.ac.uk}
\orcid{0000-0003-2925-7275}
\affiliation{%
  \institution{EPCC at the University of Edinburgh}
  \streetaddress{The Bayes Centre, 47 Potterrow}
  \city{Edinburgh}
  \country{UK}
}
\author{Maurice Jamieson}
\orcid{0000-0003-1626-4871}
\affiliation{%
  \institution{EPCC at the University of Edinburgh}
  \streetaddress{The Bayes Centre, 47 Potterrow}
  \city{Edinburgh}
  \country{UK}
}

\author{Anton Lydike}
\orcid{0009-0001-9389-8512}
\affiliation{%
  \institution{The School of Informatics, University of Edinburgh}
  \streetaddress{Informatics Forum}
  \city{Edinburgh}
  \country{UK}}
\author{Emilien Bauer}
\orcid{0009-0006-8028-3064}
\affiliation{%
  \institution{The School of Informatics, University of Edinburgh}
  \streetaddress{Informatics Forum}
  \city{Edinburgh}
  \country{UK}}
\author{Tobias Grosser}
\orcid{0000-0003-3874-6003}
\affiliation{%
  \institution{The School of Informatics, University of Edinburgh}
  \streetaddress{Informatics Forum}
  \city{Edinburgh}
  \country{UK}}


\begin{abstract}
MLIR has become popular since it was open sourced in 2019. A sub-project of LLVM, the flexibility provided by MLIR to represent Intermediate Representations (IR) as dialects at different abstraction levels, to mix these, and to leverage transformations between dialects provides opportunities for automated program optimisation and parallelisation. In addition to general purpose compilers built upon MLIR, domain specific abstractions have also been developed. 

In this paper we explore complimenting the Flang MLIR general purpose compiler by combining with the domain specific Open Earth Compiler’s MLIR stencil dialect. Developing transformations to discover and extracts stencils from Fortran, this specialisation delivers between a 2- and 10-times performance improvement for our benchmarks on a Cray supercomputer compared to using Flang alone. Furthermore, by leveraging existing MLIR transformations we develop an auto-parallelisation approach targeting multi-threaded and distributed memory parallelism, and optimised execution on GPUs, without any modifications to the serial Fortran source code.
\end{abstract}




\maketitle

\pagestyle{plain}
\copyrightnotice
\section{Introduction}
Since it was first released open source by Google in 2019 and then merged into the LLVM codebase as a sub-project thereafter, MLIR \cite{lattner2021mlir} has gained significant popularity. Enabling hierarchies of Intermediate Representations (IR) to be expressed and mixed in a structured manner, and for lowerings between these abstraction levels to be provided, numerous tools and technologies have been developed that leverage MLIR in their flow. 

One such class of tools that can strongly benefit from MLIR is that of Domain Specific Languages (DSLs), such as \cite{lange2016devito}, \cite{mudalige2019large}, \cite{clement2018claw}, which provide domain specific abstractions that enable programmes to express their problem in a high level, abstract, fashion. It is possible, using MLIR, to develop IR dialects that closely match these domain specific abstractions, and for these dialects to then be lowered to more general purpose abstractions which themselves benefit from existing lowerings to LLVM-IR and ultimately the LLVM backends. Given the rich semantic information that can often be found in domain specific abstractions about a programmer's intentions, it is often possible for the compiler framework to drive key decisions around performance and parallelism using this high level information far more effectively than, for example, working with the lower-level LLVM-IR directly. 

General purpose compilers, such as Flang \cite{flang} for Fortran, Polygeist \cite{moses2021polygeist} for C++, and Pylir for Python \cite{pylir}, have been developed which sit atop the MLIR ecosystem. Whilst these leverage the central ideas of MLIR, these general purpose language compilers do not tend to fully exploit the domain specific abstractions that are found in some MLIR dialects.

In this paper we explore the potential of combing MLIR-based general purpose compilers with domain specific MLIR abstractions by enriching the Flang compiler with the MLIR stencil dialect from the Open Earth Compiler \cite{gysi2021domain}. By automatically identifying and translating appropriate Fortran constructs into the stencil dialect during compilation, we aim to explore whether improved performance and new capabilities can be unlocked by leveraging this domain specific specialisation built atop MLIR. This paper is structured as follows; Section \ref{sec:bg} describes the background to this work, exploring the central building blocks such as MLIR, the stencil dialect, Flang and xDSL that we use in this work. Our stencil-based Flang compilation flow is then described in Section \ref{sec:stencil-optimisation}, where we highlight our approach to Fortran code stencil identification, explore how existing MLIR passes can be leveraged by our approach to target a variety of architectures and work around some of the challenges present in Flang. Performance of our approach is compared against that of using Flang alone and the Cray compiler (for CPUs) and Nvidia HPC SDK (for GPUs) in Section \ref{sec:results} across single CPU core, multi-threaded and distributed memory parallelism, and GPUs. Section \ref{sec:related_work} then surveys how our contribution relates to existing work, before Section \ref{sec:conc} draws conclusions and discusses further work.

The contributions of this paper are as follows:

\begin{itemize}
\item We demonstrate that, by exploiting stencil-based domain specific information one can deliver improved performance compared to using general purpose Flang alone for stencil codes.
\item Illustrate that domain specific MLIR abstractions and the wider ecosystem can enable the targeting of multiple HPC architectures and scales of parallelism in an automated fashion without requiring source code modifications.
\item Providing a wider study around the benefits that the composability of MLIR dialects can provide to compliment general purpose compilers for high performance workloads.
\end{itemize}

\section{Background}
\label{sec:bg}

At its core LLVM provides numerous language frontends and architecture specific backends which are connected via LLVM-IR. An LLVM frontend, such as Clang, generating LLVM-IR can therefore target any backend, and backends for a wide range of architectures including CPUs, GPUs, and FPGAs have been developed. However, LLVM-IR itself is low level, requiring significant work by the frontends in lowering to this level and resulting in potential redundancies between them.

MLIR aims to address this issue by providing a series of IR dialects and transformations between these, so that frontends can instead translate to more suitable, higher level intermediate representations. MLIR is also a framework where developers can add their own dialects and transformations, and dialects can be mixed and manipulated at different levels of abstraction, enabling progressive lowering of the abstraction level to LLVM-IR. Much of this lowering is undertaken by existing dialects and transformations, thus significantly reducing the overall software effort in developing compilers by promoting reuse between them.

MLIR has become popular since it became a sub-project of LLVM, and has the potential to revolutionise compiler development. Providing many dialects as standard, such as \emph{arith} for arithmetic operations, \emph{scf} for structured control flow which provides serial and parallel loops, \emph{memref} for memory management and data access, \emph{openmp} for OpenMP parallelism, \emph{gpu} for GPU execution, and \emph{vector} for vectorisation. Transformations exist which will manipulate dialects and lower between them. For instance, there are lowerings from the dialects listed above to the llvm dialect which corresponds to LLVM-IR. Once lowered, it is possible to generate LLVM-IR from the llvm dialect and for this to be provided to the LLVM backends.

\subsection{xDSL}
Arguably one of the challenges faced by MLIR is the steep learning curve associated with the technology. Requiring the developer to leverage C++, understand LLVM concepts, and work with the Tablegen format in order to describe dialects raises the overhead involved in development.

By contrast, xDSL \cite{xdsl} is a Python based compiler design toolkit which is 1-1 compatible with MLIR. Providing not only the majority of MLIR dialects, but also numerous additional experimental dialects too, these are expressed in IRDL \cite{fehr2022irdl} format within Python classes. xDSL enables a rapid exploration and prototyping of MLIR dialects and concepts, with a view to then committing the mature dialects and transformations into the main MLIR codebase once the concepts are proven. As xDSL is 1-1 compatible with MLIR, one is able to arbitrarily go between the two technologies during compilation.

In addition to providing many of the standard MLIR dialects \footnote{Providing a complete set of MLIR dialects is planned by the xDSL developers}, xDSL also provides additional dialects such as Distributed Memory Parallelism (DMP) which, in a technology agnostic manner, expresses parallelism across nodes. DMP can then be lowered to the MPI xDSL dialect, which specialises DMP to leverage MPI which is then lowered to interaction with the MPI library via the func dialect. 

\subsection{Stencil dialect}
\begin{lstfloat}
\begin{lstlisting}[language=Fortran, frame=lines, label=lst:fortran_stencil_example, numbers=left, caption=Sketch of Fortran stencil code example which averages all neighbouring values across a grid]
do i = 2, 255
  do j = 2, 255
    data(j,i) = (data(j,i-1)+data(j,i+1)+data(j-1,i)+data(j+1,i)) * 0.25
  enddo
enddo
\end{lstlisting}
\end{lstfloat}

A dialect provided by xDSL is the \emph{stencil} dialect which was initially developed by ETH Zurich as part of the Open Earth Compiler \cite{gysi2021domain}. A stencil is a geometric arrangement of a group of neighbouring grid cells that, by using a numerical approximation routine, relate to a specific grid cell of interest. The stencil dialect expresses stencil calculations such as that illustrated in Listing \ref{lst:fortran_stencil_example} which is calculating an average of values across neighbouring grid cells in two dimensions. Driven by nested loops, two in Listing \ref{lst:fortran_stencil_example}, the loop variables are used as array indices typically with some offset applied such as \emph{data(j,i-1)} which accesses the grid cell at the same location in the first dimension and one step before in the second dimension (arrays index precedence is from left to right in Fortran).

\begin{lstfloat*}
\begin{lstlisting}[language=mlir, frame=lines, label=lst:mlir_ssa_stencil, numbers=left, caption=Sketch of corresponding SSA-based IR leveraging the stencil dialect to represent the stencil calculation of Listing \ref{lst:fortran_stencil_example}]
%result = "stencil.apply"(%18) ({
  ^0(%data : !stencil.temp<[-1,255]x[-1,255]xf64>):
    %c0 = arith.constant 2.500000e-01 : f64
    %d0 = "stencil.access"(%data) {"offset" = #stencil.index<0, -1>} : (!stencil.temp<[-1,255]x[-1,255]xf64>) -> f64
    %d1 = "stencil.access"(%data) {"offset" = #stencil.index<0, 1>} : (!stencil.temp<[-1,255]x[-1,255]xf64>) -> f64
    %d2 = "stencil.access"(%data) {"offset" = #stencil.index<-1, 0>} : (!stencil.temp<[-1,255]x[-1,255]xf64>) -> f64
    %d3 = "stencil.access"(%data) {"offset" = #stencil.index<1, 0>} : (!stencil.temp<[-1,255]x[-1,255]xf64>) -> f64
    %t0 = arith.addf %d3, %d2 : f64
    %t1 = arith.addf %t0, %d1 : f64
    %t2 = arith.addf %t1, %d0 : f64
    %t3 = arith.mulf %t2, %c0 : f64
    "stencil.return"(%t3) : (f64) -> ()
}) : (!stencil.temp<[-1,255]x[-1,255]xf64>) -> !stencil.temp<[0,254]x[0,254]xf64>
\end{lstlisting}
\end{lstfloat*}

Stencil-based calculations are extremely common in computational simulation codes, for instance to solve systems of PDEs, \cite{brown2020highly} \cite{vazquez2014alya} \cite{de2015modeling}, and Listing \ref{lst:mlir_ssa_stencil} sketches the corresponding IR in Static Single Assignment (SSA) form using the stencil dialect for the calculation of Listing \ref{lst:fortran_stencil_example}. It can be seen that the nested loops at lines 1 and 2 of Listing \ref{lst:fortran_stencil_example} have been transformed into the \emph{stencil.apply} operator at line 1 of Listing \ref{lst:mlir_ssa_stencil} which accepts a \emph{stencil.temp} field as an input argument. There are additional stencil operators to convert from a memref to this \emph{stencil.temp} type, but these have been omitted for brevity. The \emph{stencil.access} operations at lines 4 to 7 of Listing \ref{lst:mlir_ssa_stencil} correspond to accesses on the \emph{data} array at line 3 of Listing \ref{lst:fortran_stencil_example} and each of these operations loads a specific neighbouring grid cell. Lines 8 to 11 then use the standard arith dialect to undertake the calculation which is returned from the stencil apply operator block at line 12. It is important to highlight that this \emph{stencil.apply} operator is running across the entire grid, whose lower and upper bounds are determined by the types of the input and output fields, effectively executing lines 3 to 12 for every grid cell.

It should be stressed that the code provided in this section is just an example and there is no existing lowering from Fortran to the stencil dialect. Indeed it is the objective of the work reported in this paper to be able to generate SSA such as that of Listing \ref{lst:mlir_ssa_stencil} from Fortran code. Our hypothesis is that, by transforming and simplifying applicable loops and their calculations into the stencil dialect, we unlock the potential for richer and more complex optimisations by the compiler framework. Driven by existing transformations, some of which bespoke for the stencil dialect and others standard MLIR passes, it is then possible to target different architectures such as CPUs and GPUs, as well as implicit shared memory and distributed memory parallelism in an efficient manner.

\subsection{Flang}
\label{sec:bg_flang}
Flang is the LLVM Fortran frontend, and whilst there was a previous Flang compiler for several years, known as classic Flang, this was never an official LLVM project and has been recently replaced with a ground-up rewrite built on-top of MLIR. This new Flang compiler is an official component of LLVM, and the objective is to support the full range of standard Fortran, including future versions of the language. However, at the time of writing support for Fortran at or beyond 2003 is still work in progress and yet to reach full maturity.

After lexing and parsing of Fortran code, this is then lowered to the Fortran IR (FIR) \cite{fir} dialect \footnote{There is an HL-FIR dialect in development which provides higher level FIR operations although at the time of writing the default compilation flow is still via FIR} which provides IR level expression of Fortran constructs and concepts. However, the developers of Flang have adopted a surprising design decision where FIR is then lowered directly to LLVM-IR by Flang, rather than lowering first to more general MLIR dialects such as scf and then leveraging existing MLIR passes to optimise these and ultimately lower to the llvm dialect.  

Furthermore, only a small subset of the standard MLIR dialects are registered in Flang, and conversely \emph{mlir-opt} is unaware of the FIR dialect. Consequently interoperability between FIR and many of the standard dialects is limited, which was a challenge this work and is explored further in Section \ref{sec:stencil-optimisation}.

\section{Combining Flang with stencil dialect abstractions}
\label{sec:stencil-optimisation}
\begin{figure*}
\centering
 \includegraphics[width=\textwidth]{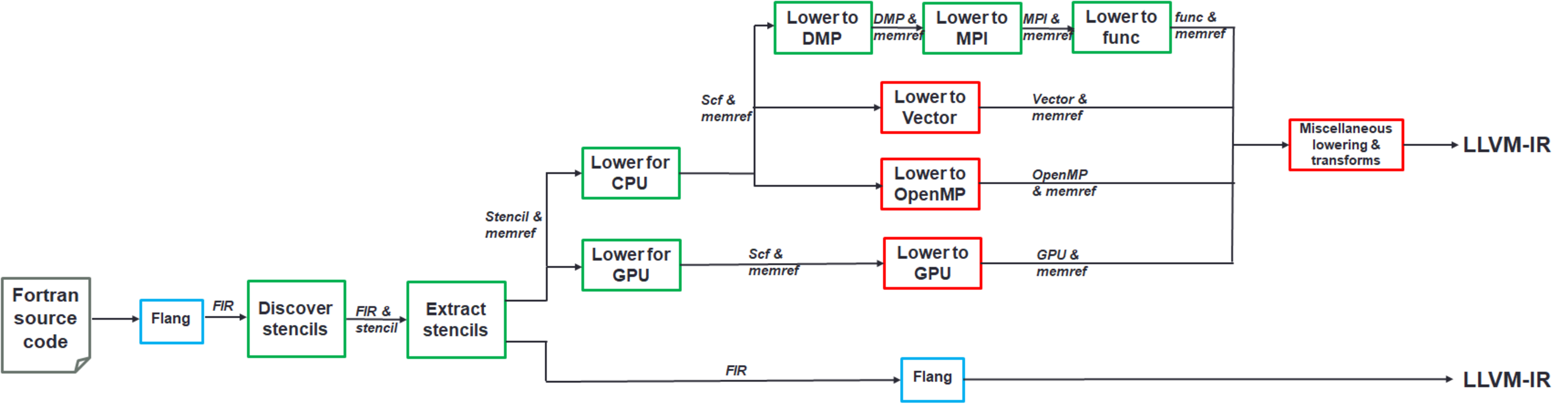}
\caption{Architectural pipeline diagram of our approach, identifying and extracting stencils from Fortran code, by operating on the Flang generated FIR, and then leveraging appropriate transformations. Boxes represent transformations and annotated arrows the dialects. Blue boxes are activities in Flang, Green boxes are transformation passes executing in xDSL, and red boxes are passes running inside MLIR.}	
\label{fig:architecture}
\end{figure*}

Figure \ref{fig:architecture} illustrates the overall architecture of our approach, where Fortran source code is processed by Flang and, using the \emph{-fc1 -emit-mlir} flags, we output the corresponding SSA-based IR in the FIR MLIR dialect. 

The SSA-based IR is loaded by xDSL which provides the FIR dialect, and our \emph{discover stencils} transformation is then run on this FIR IR to identify any stencil calculations that are being undertaken. Listing \ref{lst:sketch_algorithm} sketches the algorithm used by this pass, where all the loops in the code are first gathered and then all FIR \emph{store} operations in the module are identified and iterated over. It is first determined, for a store operation, whether it is indexed by loops in the program (line 5 of Listing \ref{lst:sketch_algorithm}) and this involves walking the IR to find the corresponding \emph{fir.coordinate\_of} operation which provides the indices into the array. The expressions which contribute to each index of \emph{fir.coordinate\_of} operation are then walked to extract the calculations involved, for instance the array access \emph{data(i,j)} of the example in Listing \ref{lst:fortran_stencil_example} will identify that the first dimension is indexed by variable \emph{i} and the second dimension by {j}. These extracted variable indices are then compared against the gathered loops to determine whether the store is being driven by nested loops and if so \emph{is\_indexed\_by\_loops} returns true.

\begin{lstfloat*}
\begin{lstlisting}[language=Python, frame=lines, label=lst:sketch_algorithm, numbers=left, caption=Sketch of stencil discovery algorithm used in the first transformation of our pipeline of Figure \ref{fig:architecture}]
identified_stencils=[]
loops = gather_program_loops(module)

for store_op in module:
  if is_indexed_by_loops(store_op, loops):
    rhs_read_ops=get_array_read_data_ops(store_op)
    unique_array_names=get_unique_array_names(rhs_read_ops)
    load_ops=[], load_ssa_results=[]
    for name in unique_array_names:
      specific_ops, specific_ssa_results=generate_stencil_field_load(name)
      load_ops=load_ops.append(specific_ops)
      load_ssa_results=load_ssa_results.append(specific_ssa_results)
    applicable_loops=get_applicable_loops(store_op, loops)
    lb, ub=get_loop_bounds_per_dim(applicable_loops)
    stencil_apply_ops=generate_stencil_apply(load_ssa_results, lb, ub, store_op)
    store_ops=generate_stencil_store(store_op)
    identified_stencils.append((applicable_loops, load_ops+stencil_apply_ops+store_ops))

for stencil in identified_stencils:
  applicable_loops=stencil[0]
  ops=stencil[1]
  top_level_loop=find_top_level_loop(applicable_loops)
  insert_ops_before(top_level_loop, ops)

for loop in loops:
  if loop is empty:
    remove_loop(loop)

merge_stencils_if_possible(module)
\end{lstlisting}
\end{lstfloat*}

The SSA-based IR is then walked to locate array read operations which are on the right hand side of the calculation. In addition to capturing the array variable being accessed, array indices are walked in a manner similar to the store operation however in addition to capturing the index variables, this also identifies offsets where, for example, \emph{data(j,i-1)} would record \emph{j} indexing the first dimension and {i} minus 1 as the second dimension. Using this information, the list of arrays that are accessed as part of this calculation are determined at line 7 and then, for each of these, the corresponding stencil loading operations are then generated at line 10, with the operations and SSA value returned, the later then used as input to the \emph{stencil.apply} operation at line 15.

The loops which drive this stencil calculation are then identified at line 13 of Listing \ref{lst:sketch_algorithm}, and from these the lower and upper bounds in each dimension. The \emph{stencil.apply} operation is generated at line 15 and the \emph{generate\_stencil\_apply} function first creates a \emph{stencil.apply} based upon the lower and upper bounds and with each array load SSA result as an input, before then walking the SSA IR to identify the mathematical expressions and data dependencies that contribute to the stencil calculation. These mathematical expressions are removed from the original FIR code and placed into the stencil, with the data dependencies being replaced by the \emph{stencil.apply} operation which sets the field offset based upon the offsets already gathered in \emph{rhs\_read\_ops}, for instance 0 for dimension zero and minus 1 for dimension one in the example above.

Operations to store the result of applying the stencil back to the memref are generated at line 16 of Listing \ref{lst:sketch_algorithm}, before the applicable loops that this stencil involves along with the generated instructions are stored in the \emph{identified\_stencils} list. The next step is then to add these stencils into the IR, which are added directly preceding the outer most loop involved in the stencil calculation. This is handled between lines 19 and 23, where the top level loop for the stencil is identified at line 22 and the stencil operations added into the list of operations directly preceding this at line 23. 

The bodies of all FIR loops are then walked at lines 25 to 27 and those loops which are now empty, because their operations have been transformed into the stencil dialect and extracted, are removed. Lastly, a pass is undertaken across the SSA-based IR to merge any stencils that are located next to each other and share the same lower and upper bounds.

It should be noted that Listing \ref{lst:sketch_algorithm} is a sketch of the algorithm, and there are some specific complexities omitted for brevity. For instance, differences in how arrays are represented in FIR by Flang if they are stack or heap allocated mean that there are different possible routes when walking backwards from \emph{fir.store} and \emph{fir.load} operations. Other features, such as accessing loop indexes, constants and non-stencil based variables within a stencil calculation are also supported and translated to the corresponding operations in the stencil dialect. 

Once stencils have been identified, the SSA-based IR is now in the form of the FIR and stencil dialects being mixed together. However, this is problematic because, as explained in Section \ref{sec:bg_flang}, Flang is unaware of many of the standard MLIR dialects, and likewise the MLIR driver tool, \emph{mlir-opt}, is unaware of FIR. Consequently the FIR portions of the IR must be separated from that of the stencil dialect. This is performed by the \emph{extract stencil} pass which lifts the stencil components out as a function into a separate MLIR module, which will then be called from FIR by a function call. As these are then separate MLIR modules they can then be compiled by different flows, i.e. one using Flang and the other \emph{mlir-opt}, and linked together at runtime. Due to this requirement of the stencil dialect IR not containing any FIR dialect IR, during the phase in Listing \ref{lst:sketch_algorithm} where we build up the \emph{stencil.apply} operator by extracting the mathematical expressions from FIR into the stencil, we need to convert FIR operations into standard dialects. This is simplified significantly by the fact that Flang leverages the standard arith and math dialects for arithmetic and mathematical operations, which are registered with \emph{mlir-opt} and-so can be handled. However, FIR data conversions and miscellaneous operations, such as \emph{fir.no\_reassoc} which is used to prevent operator reassociation, do need to be converted into their standard MLIR dialect counterparts.

A further challenge is in providing data interoperability between FIR, which contains its own bespoke data representation operations and types, and the stencil dialect which uses the standard memref dialect. The only way to support this is to convert the FIR data to an FIR \emph{llvm\_ptr} type, and pass this to the stencil function which then builds the memref from this. In-fact FIR is entirely isolated from a typing perspective, as one can only reduce to its own, FIR dialect, representation of an \emph{llvm\_ptr}, and not the \emph{llvm\_ptr} in the llvm dialect. Furthermore, it is not possible in the FIR-based IR module to leverage an unrealized conversion cast, which would enable one to go from an FIR \emph{llvm\_ptr} to an LLVM \emph{llvm\_ptr} from a typing perspective, because the builtin dialect is not registered with Flang. However, \emph{llvm\_ptr} in the FIR and llvm dialects is semantically identical and passing an argument of type FIR \emph{llvm\_ptr} to a function that accepts LLVM \emph{llvm\_ptr} is allowed when linking the resulting object files.

As per Figure \ref{fig:architecture}, the IR module containing the stencil dialect portion is then transformed either for CPU or GPU using the xDSL stencil lowering. In fact these lowerings are the same source code, but driven by a command line option to tune for the architecture in question, for instance the CPU lowering converts the top level loop into \emph{scf.parallel} and nested inner loops into \emph{scf.for}, whereas the GPU lowering attempts to coalesce the loops into a single \emph{scf.parallel} loop. For  GPU, shared memory parallelism and single core execution the lowered IR is then transformed by existing MLIR scf transformation passes, for instance \emph{convert-scf-to-openmp} to lower to the OpenMP dialect, \emph{convert-parallel-loops-to-gpu} for GPU execution, and \emph{scf-for-loop-specialization} for vectorisation. Alternatively, if the programmer wishes to leverage distributed memory parallelism then they can apply the \emph{lower to DMP} transformation which will transform to the DMP dialect in xDSL, which can then be lowered to MPI and the corresponding function calls via two subsequent passes. 

Irrespective of the specific passes and targets, this IR is then transformed through a series of existing MLIR miscellaneous passes, such as lowering dialects such as math and memref to LLVM, reconciling unrealised conversion casts, and canonicalization. The MLIR pass pipeline for GPU target is reported in Listing \ref{lst:mlir_pipeline} and there are several aspects to highlight. Firstly, for GPUs, we found that this was very sensitive and slight modifications to the pipeline would silently fail to generate GPU target binary code and run only on the CPU. As the GPU binary is embedded in the single generated MLIR CPU file, this is very easy to miss. Secondly, there is a sensitivity to the loop tiling factors on the GPU, provided to the \emph{scf-parallel-loop-tiling} pass as arguments, as these can make both an impact on performance and furthermore some values can result in runtime failures on the GPU. We have had to find these optimal values empirically, although the values illustrated in Listing \ref{lst:mlir_pipeline} perform well across a range of kernels, and an improved MLIR pass that avoids the need for these specific numbers would be beneficial. Thirdly, several transformations by default lower to opaque pointers, whereas in our flow we favour including the type and size. Consequently, we provide the option to use transparent pointers instead.

\begin{lstfloat}
\begin{lstlisting}[language=bash, frame=lines, label=lst:mlir_pipeline, numbers=none, caption=Command issued to lower transformed stencil IR to GPU via \emph{mlir-opt}]
mlir-opt --pass-pipeline="builtin.module(test-math-algebraic-simplification,scf-parallel-loop-tiling{parallel-loop-tile-sizes=32,32,1},canonicalize,test-expand-math,func.func(gpu-map-parallel-loops), convert-parallel-loops-to-gpu, fold-memref-alias-ops,finalize-memref-to-llvm{index-bitwidth=64 use-opaque-pointers=false},lower-affine, gpu-kernel-outlining,func.func(gpu-async-region),canonicalize,convert-arith-to-llvm{index-bitwidth=64},finalize-memref-to-llvm{index-bitwidth=64 use-opaque-pointers=false},convert-scf-to-cf,convert-cf-to-llvm{index-bitwidth=64},finalize-memref-to-llvm{use-opaque-pointers=false}, gpu.module(convert-gpu-to-nvvm,reconcile-unrealized-casts,canonicalize,gpu-to-cubin),fold-memref-alias-ops,lower-affine,gpu-to-llvm{use-opaque-pointers=false},finalize-memref-to-llvm{index-bitwidth=64 use-opaque-pointers=false},reconcile-unrealized-casts)" stencil.mlir
\end{lstlisting}
\end{lstfloat}

Once the stencil has been lowered to the LLVM MLIR dialect, this is then transformed into LLVM-IR, compiled into an object file by Clang. The isolated FIR IR module is compiled by Flang into an object file and these are then linked together into an executable. In the manner described in this section we have intercepted the FIR-based IR representation of a Fortran code, transformed applicable parts into stencils and then applied specialist optimisations and transformations on these.

\section{Results and evaluation}
\label{sec:results}
\subsection{Experimental environment}
In this section we use two stencil-based codes as benchmarks, the first solving LaPlace's equation for diffusion in three dimensions via a Gauss Seidel solver. An iterative algorithm containing an outer loop operating across the entire domain and generating a progressively improving sequence approximate solutions, for each grid cell at each iteration this benchmark works with a 7-point stencil, the orthogonal neighbours in three dimensions, and averages values across the six neighbouring cells. Consequently, there are six floating point operations required per grid cell.

The second benchmark is the Piacsek and Williams advection scheme \cite{pwadvection}, commonly used by Met Office codes such as the MONC high-resolution atmospheric model \cite{brown2020highly}, for calculating the movement of quantities through the atmosphere due to kinetic effects (i.e. wind). This is more complex than the first benchmark, containing three separate stencil computations across three fields which are then fused by our stencil transformation into a single stencil region. There are 63 floating point operations required per grid cell.

CPU based benchmarks conducted in this paper have been run on ARCHER2, a Cray-EX which is the UK national supercomputer. Each node contains two 64-core AMD EPYC 7742 (Rome, Zen 2) CPUs with 256 GB of memory. There are 16 cores per NUMA region, providing a total of 8 NUMA regions per node. Nodes are interconnected via HPE Cray Slingshot, delivering two 100 Gbps bi-directional links per node. 

GPU based benchmarks are run on Cirrus, an HPE/SGI 8600 HPC system, where the GPU nodes provide Nvidia Tesla V100-SXM2-16GB (Volta) GPUs and two 20-core Intel Xeon Gold 6248 (Skylake) CPUs with 384 GB of memory. Nodes are interconnected via an FDR single infiniband (IB) fabric.

All experiments are averaged over five runs, and we use version 15 of the Cray Compilation Environment (CCE), version 16 of LLVM, version 0.14.0 of xDSL, and version 22.11 of the Nvidia HPC SDK. Performance is reported in terms of throughput, using the metric millions of grid cells per second (MCells/s). Because the Gauss Seidel benchmark is simpler, requiring only 6 floating point operations per grid cell compared to 63 FP operations for the PW advection benchmark, this will naturally deliver the higher throughput of the two benchmarks and therefore a throughput comparisons between benchmarks is less interesting than comparisons for different sizes and configurations of each individual benchmark.

\subsection{Single node CPU performance}

\begin{figure}[htb]
\centering
 \includegraphics[width=\columnwidth]{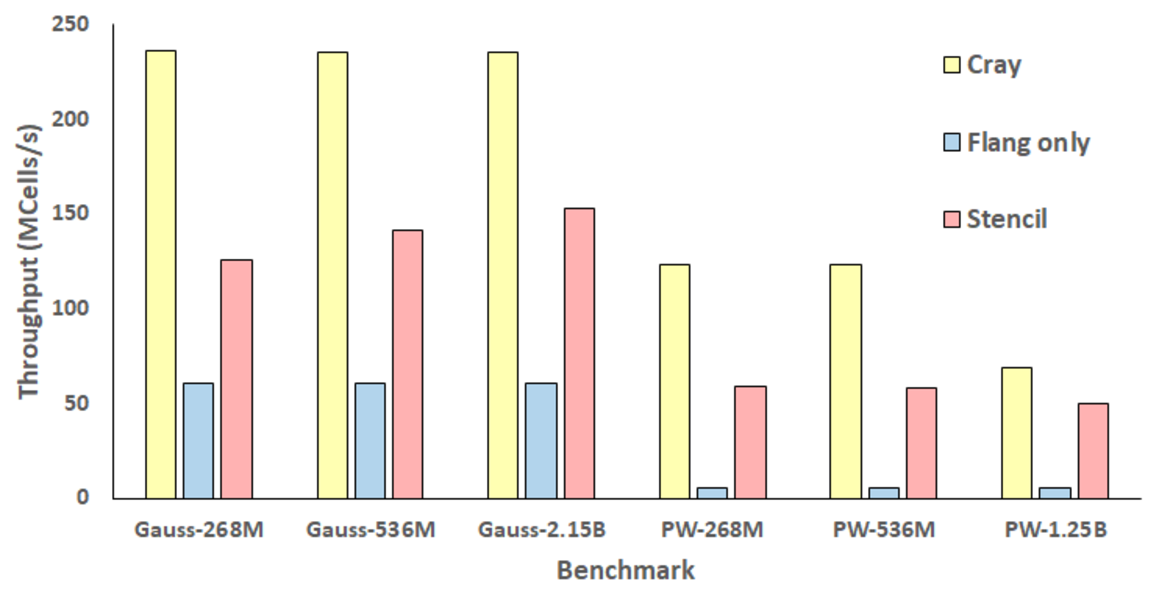}
\caption{Single core performance comparison for Gauss Seidel and PW advection benchmarks using different problem sizes across Cray, Flang only and our stencil approach.}	
\label{fig:single_core_performance}
\end{figure}

Figure \ref{fig:single_core_performance} reports performance, as throughput, achieved on a single core of ARCHER2 for our two benchmarks at different problem sizes. We are comparing approaches leveraging the Cray compiler (\emph{Cray} in Figure \ref{fig:single_core_performance}), Flang on its own (\emph{Flang only} in Figure \ref{fig:single_core_performance}), and our stencil approach (\emph{Stencil} in Figure \ref{fig:single_core_performance}) described in Section \ref{sec:stencil-optimisation}. It can be seen that the Cray compiler provides excellent performance, which is inline with previous community experience, and using Flang directly delivers the lowest performance. Our stencil approach achieves performance that lies between the Cray compiler and Flang, and when profiling we found that the executable generated by the Cray compiler undertakes considerably more vectorisation than our stencil approach, even though we use the \emph{scf-parallel-loop-specialization} transformation pass in our CPU flow. It can be seen clearly however, that by leveraging our stencil optimisation the programmer is able to obtain significantly higher performance than Flang alone, especially for the PW advection benchmark.

\begin{figure}[htb]
\centering
 \includegraphics[width=\columnwidth]{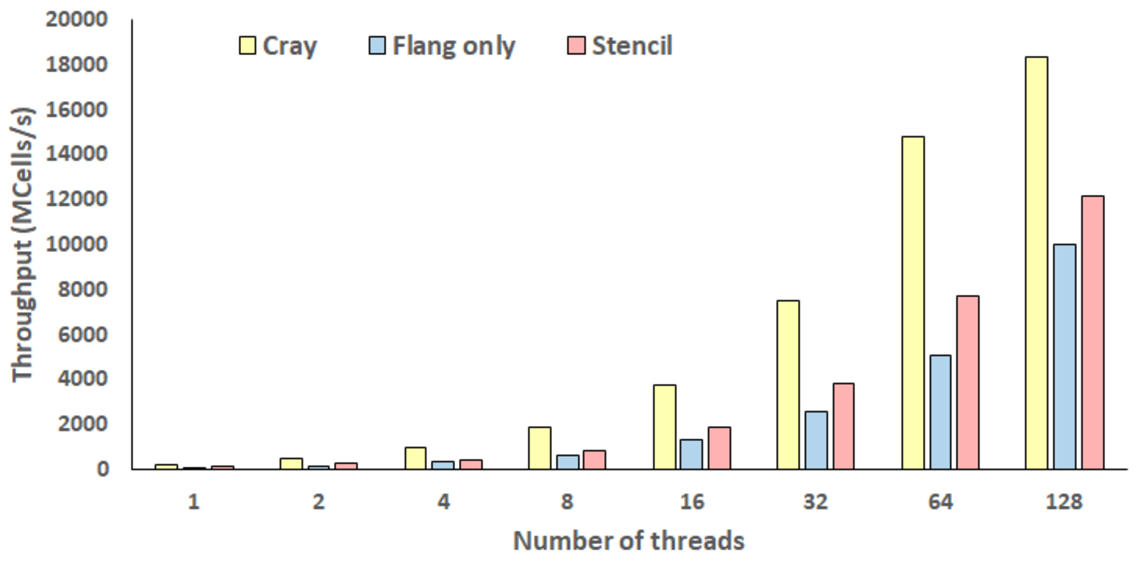}
\caption{Multithreaded performance comparison for Gauss Seidel benchmark using OpenMP with a problem size of 2.1 billion grid cells across Cray, Flang only and our stencil approach.}	
\label{fig:threading_gauss}
\end{figure}

\begin{figure}[htb]
\centering
 \includegraphics[width=\columnwidth]{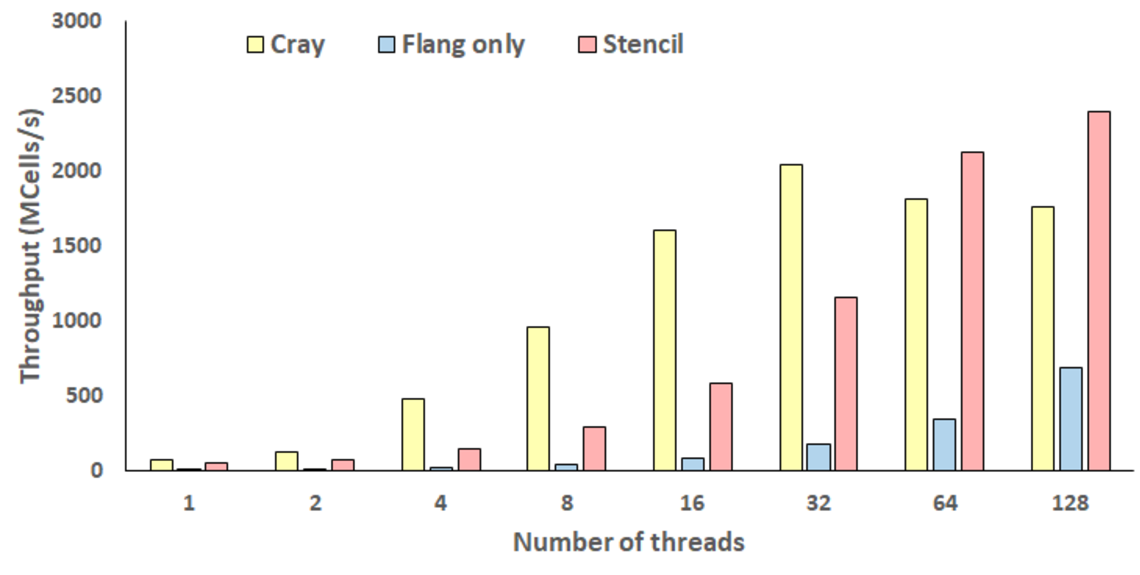}
\caption{Multithreaded performance comparison for PW advection benchmark using OpenMP with a problem size of 2.1 billion grid cells across Cray, Flang only and our stencil approach.}	
\label{fig:threading_pw}
\end{figure}

Figures \ref{fig:threading_gauss} and \ref{fig:threading_pw} report multithreaded performance for the Gauss Seidel and PW advection benchmarks respectively when run on the largest problem size of 2.1 billion grid cells. Similarly to single core performance, it can be seen that generally the Cray compiler delivers the best performance, with vanilla Flang the lowest and our stencil approach between these two. However, it can be seen in Figure \ref{fig:threading_pw} that for the PW advection benchmark our stencil approach delivers the highest performance at 64 and 128 thread. Furthermore, it should be highlighted that the Cray and vanilla Flang multithreaded experiments are being executed with hand written OpenMP code, requiring the programmer to modify their code to run multithreaded. By comparison, our stencil approach is leveraging unchanged serial code with the OpenMP parallelism added automatically by the compiler.

\subsection{GPU performance}

\begin{figure}[htb]
\centering
 \includegraphics[width=\columnwidth]{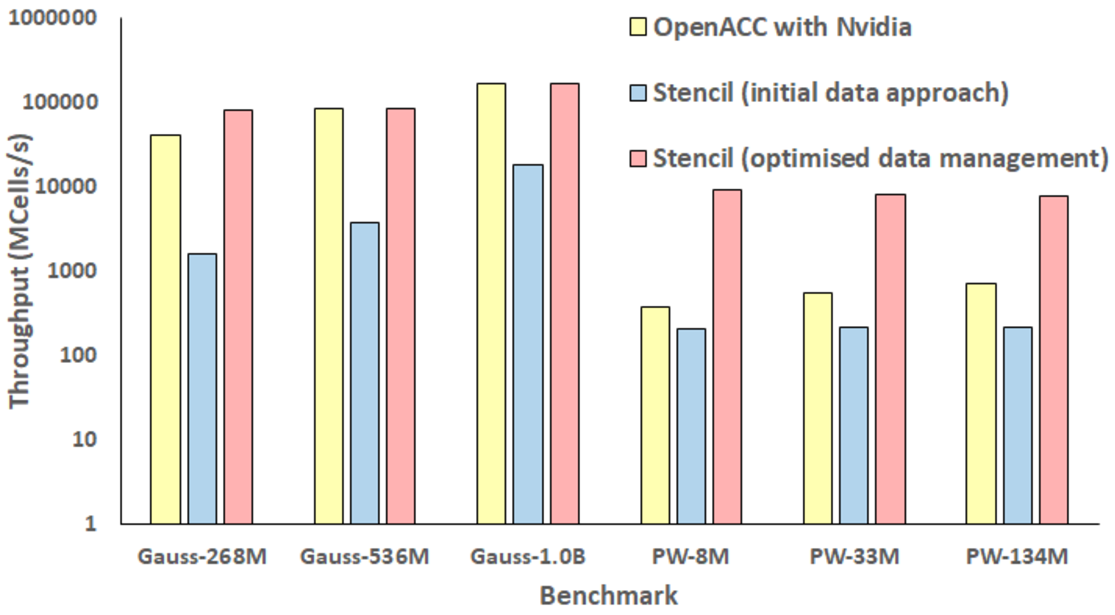}
\caption{Log scale GPU performance comparison on Nvidia V100 GPU for both benchmarks at different problem sizes. Comparing OpenACC compiled with the Nvidia compiler against our stencil approach. Stencil approach reports figures for both our initial and optimised data management strategies.}	
\label{fig:gpu_performance}
\end{figure}

Figure \ref{fig:gpu_performance} reports the log scale performance of our two benchmarks, at different problem sizes, running on a Nvidia V100 GPU. For each configuration there are three numbers reported, firstly the performance obtained from a manual porting of the code to the V100 GPU using OpenACC and compiled with the Nvidia compiler (\emph{OpenACC with Nvidia} in Figure \ref{fig:gpu_performance}). There are two results for our stencil approach where the initial approach we adopted, (\emph{Stencil (initial data approach)} in Figure \ref{fig:gpu_performance}), was for the GPU dialect to manage all data movement via the \emph{gpu.host\_register} operation on all stencil data arrays. However, we found that this delivered very poor performance and when profiling discovered it was due to excessive movement of data between the host and GPU over PCI express. Effectively, this \emph{gpu.host\_register} operation is allocating data on the host and moves it across on demand, without effective caching which was causing significant runtime overhead.

Consequently, we developed our own approach, (\emph{Stencil (optimised data approach)} in Figure \ref{fig:gpu_performance}), to managing the memory by developing a bespoke transformation pass. This walks the SSA-based IR just after the stencil extraction pass described in Section \ref{sec:stencil-optimisation}, and identifies what data must be placed on the GPU and when. Additional functions are added into the extracted stencil module to call operations in the GPU dialect for data allocation, movement, and deallocation and these are called from FIR using the same approach as when calling stencil execution. The FIR IR holds references to the GPU allocated data as FIR LLVM pointers which are provided to the stencil execution functions as arguments and can also be used, in combination with the FIR data reference, to copy data between the host and device. It was found that this transformation was highly effective and by undertaking this direct management of data via a transformation pass we were able to achieve significant performance as can be seen in the log scale results of Figure \ref{fig:gpu_performance}.

It can be seen in Figure \ref{fig:gpu_performance} that the data optimised version of our stencil approach is very competitive against the hand written OpenACC compiled with the Nvidia compiler. For the Gauss Seidel benchmark our approach outperforms the Nvidia compiler with hand written OpenACC for the smallest problem size and is comparable for the two larger problem sizes. Whilst there is considerably more work per grid cell, and hence a lower throughput in terms of grid cells processed per second, our optimised approach out performs the manually written OpenACC for all sizes with the PW advection benchmark. When profiling the OpenACC code we found that, due to the use of unified memory, there were numerous data access stalls. The overhead is far less than that of the MLIR managed data approach, but it does still add considerable overhead. Whilst it would likely be possible to obtain increased performance with the manual code by controlling all the data movement directly in code, this would further increase the complexity. By contrast, in our approach the programmer's Fortran source code is unchanged to run on the GPU.

\subsection{Distributed memory performance}

As described in Section \ref{sec:stencil-optimisation}, there is also a lowering in xDSL from the stencil dialect to MPI via an intermediate Distributed Memory Parallelism (DMP) dialect. Whilst this is somewhat experimental, it is worth exploring the performance that we can obtain when running on an unmodified Fortran code compared to a hand-crafted MPI parallelisation to explore the potential for automatic distributed memory parallelisation of serial code. Of our two benchmarks, we selected Gauss Seidel only for this experiment because it operates in iterations, requiring a halo swap between iterations. The PW advection benchmark by comparison is a kernel called from a larger code base which then undertakes halo swapping before the next timestep when other calculations have completed.

\begin{figure}[htb]
\centering
 \includegraphics[width=\columnwidth]{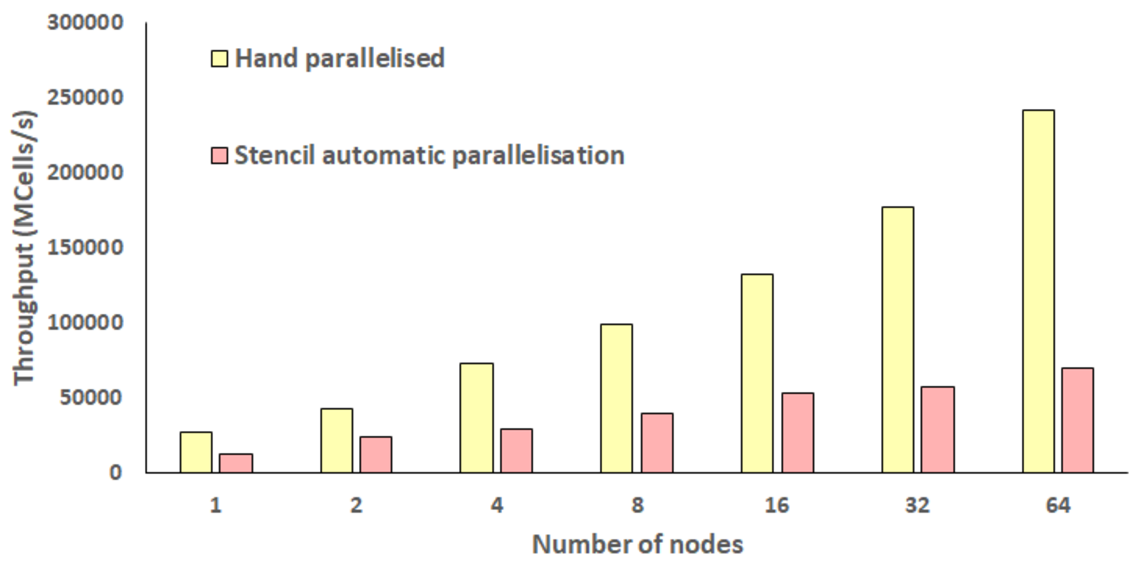}
\caption{Distributed memory performance of Gauss Seidel benchmark across nodes of ARCHER2, using one MPI process per core (128 cores per node). Global problem size of 17 billion grid cells comparing a hand-parallelised version compiled with the Cray compiler to the distributed memory version auto-generated from our flow.}	
\label{fig:dist_mem_performance}
\end{figure}

Figure \ref{fig:dist_mem_performance} reports the distributed memory performance of the Gauss Seidel benchmark running on ARCHER2, where there are 128 cores per node and we are mapping one MPI process to each core. We decompose the 3D space into two dimensions, and the results compare a hand parallelised version with the Cray compiler, (\emph{Hand parallelised} in Figure \ref{fig:dist_mem_performance}),  against the automatic parallelisation obtained via lowering our stencil dialect IR to the xDSL MPI dialect via the DMP dialect (\emph{Stencil automatic parallelisation} in Figure \ref{fig:dist_mem_performance}). It can be seen that the hand parallelised version outperforms our automatically parallelised version which is for two reasons. Firstly, as reported for single core runs, the Cray compiler is very good at vectorisation and delivering high single core performance which is relevant for the baseline performance. Secondly, the hand parallelised version scales better than the automatic parallelisation via the DMP dialect.

Whilst this demonstrates that there is still work to be done in the DMP and MPI xDSL dialects to obtain optimal performance, it should be noted that we are still able to take an unmodified Fortran serial code and for it to run across 8192 CPU cores and obtain a throughput of around 70,000 million grid cells per second. We believe that this demonstrates the potential of our approach, even if the individual components still need further enhancement to reach a level of maturity where they can match the performance of hand parallelised code.

\section{Related work}
\label{sec:related_work}
There are several existing frameworks and DSLs which aim to optimise the execution of stencil-based computations. For instance, the Pochoir \cite{tang2011pochoir} stencil compiler enables programmers to write their code in C++ and use bespoke templates to drive a higher level description of their stencil computation. This is then provided to the Pochoir compiler which undertakes source to source translation as a preprocessing step, the results of which are then compiled by the Intel compiler. Leveraging Cilk \cite{blumofe1995cilk}, which provides multithreading parallelism capabilities, their source to source translation tool generates Cilk based code from the programmer's source code. 

ExaStencils \cite{lengauer2020exastencils} is a stencil-based framework, and this supports a wider range of execution targets than Pochoir. Similar to our approach, ExaStencils supports shared and distributed memory parallelism as well as execution on GPUs. However, similarly to Pochoir users must learn new abstractions and explicitly port their code to this technology, whereas our approach leverages unmodified serial Fortran code. The major disadvantage with both Pochoi and ExaStencils is the siloed nature of their compilation stacks, where these have been developed as bespoke compilers and share no underlying infrastructure with other projects. This is especially important given the effort undertaken to optimise stencil execution by these projects, which not only requires significant investments of time to initially develop, but then a continuing maintenance burden to support new architectures and fix bugs. Adoping DSLs like these results in a risk for users, as they can not be sure about the long term future of these technologies, especially as many are developed during research projects with a fixed end date. 

By contrast, our approach leverages the popular LLVM/MLIR ecosystem which is actively developed and maintained by a large community including individuals, academic organisations, and commercial companies. Furthermore, the bespoke nature of our contribution is a small part in the overall compilation flow where, as described in Section \ref{sec:stencil-optimisation}, we provide stencil discovery and extraction transformations for FIR, with the generated SSA-based IR then transformed by passes developed by other groups, many of which are in the main MLIR codebase. Consequently, programmers can have much more confidence leveraging an approach that sits within the MLIR ecosystem because of the significant investment that has, and i,s being made into this project.

LLVM is used extensively in HPC as a growing number of compilers from vendors are now built upon the framework. Prominent HPC compiler teams, such as those at Cray, Intel, ARM, AMD and Nvidia have made significant investments in LLVM and MLIR, and many of these organisations produce products build upon the technologies. There are additional community activities exploring optimisation opportunities provided by LLVM and MLIR, for instance \cite{essadki2023code} which explored the use of MLIR to optimise the execution of stencil-based codes. Developing their own \emph{cfd} dialect which contained a stencil operation, although this is significantly more limited than the Open Earth Compiler's stencil dialect used in this work, the authors were able to demonstrate that the MLIR ecosystem is beneficial for optimising these calculations in a multi-threaded environment. However they did not integrate their work with existing, general purpose, MLIR compilers or attempt to undertake the stencil discovery and extraction from code as explored in this paper. Furthermore, by only considering multi-threaded CPU execution of their dialect they limit the benefits to be gained from the MLIR ecosystem such as GPU execution.

Lastly, DaCE \cite{ziogas2021productivity} is a parallel programming framework that operates over codes typically written in Python and converts these into a dataflow Directed Acyclical Graph (DAG) IR representation. This is then used by DaCE to generate optimal code for a variety of targets including shared and distributed memory CPUs and GPUs. Providing interoperability with MLIR via the DaCE dialect \cite{ben2023bridging}, the ability to move between DaCE dataflow and MLIR control flow representations can deliver improved executable performance. Furthermore, Stencilflow \cite{de2021stencilflow} provides the ability to translate stencils into the DaCE dataflow DAG IR with the primary objective of enabling these to be efficiently executed on FPGAs. However, at the time of writing, Stencilflow is not actively maintained, and furthermore does not undertake the automated discovery and extraction of stencils from existing code as described in this paper. It would also likely be impractical to integrate DaCE with MLIR-based compilers, such as Flang, due to its extensive nature and the additional maintenance dependency. By contrast, our approach is far more lightweight and could be integrated by the addition of our two additional compiler passes into these tools. 

\section{Conclusions and further work}
\label{sec:conc}
In this paper we have explored the potential to enrich general purpose compiler flows by leveraging domain specific information via MLIR dialects and transformations. By discovering and extracting stencils from the FIR intermediate representation generated by Flang, we have demonstrated that it is possible to then exploit this domain specific information and transformations to target numerous architectures and forms of parallelism. Flang provides an unusual approach to leveraging MLIR, by translating the FIR dialect directly into LLVM-IR rather than into the existing standard MLIR dialects which themselves are lowered. This does decrease composability between dialects, but as demonstrated in this paper can be worked around by extracting dialects that are not registered with Flang into separate modules and compiling independently. 

Our main objective was to explore whether a fusion of domain specific abstraction with the general purpose Flang compiler would enable us to deliver improved performance compared with Flang alone and, using two benchmarks, we demonstrated that our approach delivers around a two time speed up for the iterative Gauss Seidel solver and approximately a 10 times speed up for the atmospheric PW advection scheme on a single CPU compared to Flang. We found that the Cray compiler delivers impressive performance, considerably outperforming Flang and our stencil approach. However, it should be borne in mind that the Cray compiler is a mature product and the result of many years of development, furthermore it is not available for general release and can only be found on Cray supercomputers. Nevertheless, our stencil approach did outperform the Cray compiler for the PW advection benchmark when run shared memory at 64 and 128 threads. Consequently we have demonstrated that to help close the performance gap between Flang and compilers such as the Cray compiler, then leveraging other existing parts of the MLIR ecosystem can be beneficial.

On a V100 GPU our stencil flow, automatically porting the Fortran code to GPUs from the programmer's perspective, delivered similar performance to a hand written OpenACC implementation for the Gauss Seidel benchmark, and our approach was on average approximately 15 times faster than the hand written OpenACC for the PW advection benchmark. Furthermore, we explored lowering our stencils via the existing xDSL distributed memory and MPI dialects on up to 8192 cores of ARCHER2. Whilst hand parallelised distributed memory code did perform and scale better than the automatic parallelised version, the fact that we were able to scale to 8192 cores is encouraging given that the same, unchanged, Fortran source code was used for this experiment as for the single CPU, multi-threaded CPU, and GPU experiments.

There are five avenues that would be interesting to explore as further work. Firstly, single CPU performance optimisations in the scf dialect lowering, especially targeting improved vectorisation, would be worthwhile. Secondly, the xDSL DMP and MPI dialects are currently in active development and the results of this paper demonstrate that whilst they are currently usable they would benefit from additional optimisation and tuning to match the performance of existing hand crafted codes. Thirdly, we believe that it would be useful to pursue combining the stencil optimisation reported in this paper with general purpose MLIR compilers so that they can benefit from this optimisation. As this work has been developed with Fortran as the focus, then Flang integration would likely be the easiest to achieve, although we would need to consider whether to integrate with FIR or wait for the new HL-FIR to be fully matured. However our central algorithm and associated transformations could also be adapted to other languages and benefit a wider set of MLIR based compilers such as Polygeist and Pylir. Fourthly, we believe that it would be worth exploring the potential of lowering FIR into the standard MLIR dialects rather then directly to LLVM-IR. This could reduce the maintenance burden as lowering to LLVM-IR from the standard dialects would exploit more shared passes, and furthermore would also aid in bringing additional dialects into the Flang ecosystem. Lastly, combining distributed memory parallelism with GPU execution, enabling multi-node GPU execution, potentially in combination with a communication technology such as NVLink \cite{foley2017ultra}, would be worthwhile.

We conclude that this study highlights that there is a significant potential for MLIR to enrich our existing programming languages and development, not only by leveraging domain specific knowledge to deliver improved performance, but to also potentially support delivering new capabilities for codes such as automatic parallelisation and architecture portability. Using the existing building blocks of MLIR, we have found that it can be very effective to develop bespoke components and then integrate these with the existing ecosystem.

\begin{acks}
This work has been funded by the xDSL ExCALIBUR EPSRC project. This work used the ARCHER2 UK National Supercomputing Service (https://www.archer2.ac.uk). This work used the Cirrus UK National Tier-2 HPC Service at EPCC (http://www.cirrus.ac.uk) funded by the University of Edinburgh and EPSRC (EP/P020267/1)
\end{acks}

\bibliographystyle{ACM-Reference-Format}
\bibliography{sample-base}


\end{document}